# Security in Distributed Storage Systems by Communicating a Logarithmic Number of Bits

Theodoros K. Dikaliotis

Department of Electrical Engineering

California Institute of Technology

Email: tdikal@caltech.edu

Alexandros G. Dimakis

Department of Electrical Engineering

University of Southern California

Email: dimakis@usc.edu

Tracey Ho

Department of Electrical Engineering

California Institute of Technology

Email: tho@caltech.edu

**Abstract**

We investigate the problem of maintaining an encoded distributed storage system when some nodes contain adversarial errors. Using the error-correction capabilities that are built into the existing redundancy of the system, we propose a simple linear hashing scheme to detect errors in the storage nodes. Our main result is that for storing a data object of total size $\mathcal{M}$ using an $(n,k)$ MDS code over a finite field $\mathbb{F}_q$, up to $t_1 = \lfloor (n-k)/2 \rfloor$ errors can be detected, with probability of failure smaller than $1/\mathcal{M}$, by communicating only $O(n(n-k)\log \mathcal{M})$ bits to a trusted verifier. Our result constructs small projections of the data that preserve the errors with high probability and builds on a pseudorandom generator that fools linear functions. The transmission rate achieved by our scheme is asymptotically equal to the min-cut capacity between the source and any receiver.

## I. INTRODUCTION

We study the security and data integrity of distributed storage systems that use coding for redundancy. It is well known that maximum distance separable (MDS) codes can offer maximum reliability for a given storage overhead and can be used for distributed storage in data centers and peer-to-peer storage systems like OceanStore [1], Total Recall [2], and FS2You [3], that use nodes across the Internet for distributed file storage and sharing. In this paper we are interested in dealing with errors in the encoded representation. The errors could be introduced either through (unlikely) hard drive undetected failures or through a malicious or compromised server in the storage network.

This second threat is much more eminent when the system uses network coding to maintain the redundancy of the encoded system as proposed recently [4]. To illustrate this consider a large data object that has size $\mathcal{M}$ bits. If this object is to be stored on $n$ servers, depending on the desired redundancy, an $(n,k)$ linear MDS code can be used, dividing the object into $k$ packets of size $\mathcal{M}/k$ each, and storing an encoded packet at each server. Assuming the code is over a finite field $\mathbb{F}_q$, requiring $\log q$ bits to represent each symbol, each server will also need to keep a header denoting the coding coefficients of the linear combinations stored on the server (see section II for the details) and the size of this header is larger than the size of the useful data if the code is used only once. For this reason it was proposed that the same code is used several times [5] by dividing each packet into $N$ symbols of $\log q$ bits and repeating the same code $N$ times. If $N >> n$ the overhead of storing the coefficients becomes negligible. We refer to this as the $N$–extended version of an MDS code, shown in Figure 2 for the $(4, 2)$ code used in Figure 1.

[0]This work has been supported partially by NSF grant CNS-0905615, partially by the Air Force Office of Scientific Research under grant FA9550-10-1-0166 and Caltech's Lee Center for Advanced Networking

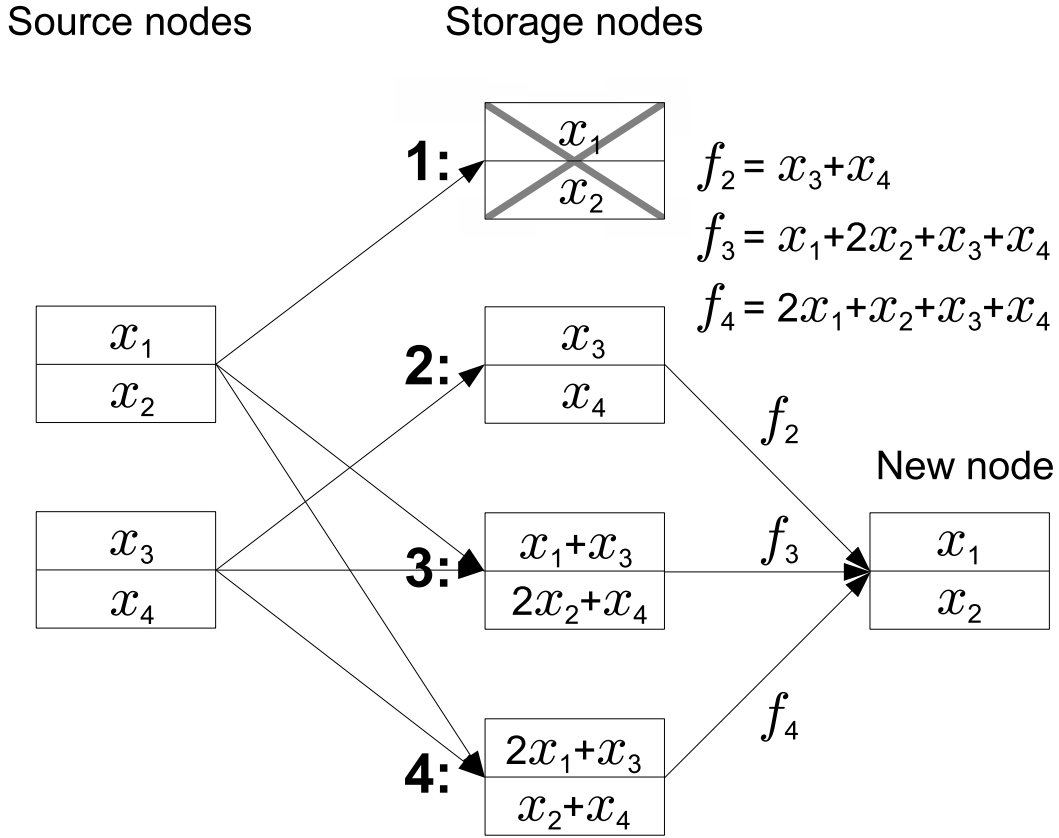

Fig. 1. A $(4, 2)$ MDS code along with the repair of the first storage node. Each node stores two packets and any two nodes contain enough information to recover all four data packets. In this example the first node leaves the system and a new node is formed by communicating linear combinations $f_2, f_3, f_4$ which can be used to solve for $x_1, x_2$ at the new node.

Observe that in this example, each node is storing two linear combinations, (rows) as opposed to one. This *sub-packetization* is performed to facilitate *repair* through network coding as proposed in [4]. The problem of repair consists of constructing a new encoded node by accessing as little information from existing encoded nodes. In the example of Figure 1, we assume that the first storage node failed and the redundancy of the system needs to be refreshed. This is achieved by communicating "small" linear combinations $f_2, f_3, f_4$ of the encoded packets from nodes $2, 3$, and $4$ each of size $1/2$ of what each node is storing, which as proven in [4], is information theoretically minimal. As storage nodes leave the system and new ones are added, this forms a dynamic storage network that keeps a fixed redundancy and reliability by building new encoded packets from already existing ones. The problem of security should now be clear: even if a single node in this storage system is compromised and participates in this repair process, then it can send incorrect linear combinations that will create erroneous packets at the new nodes. All new nodes using these linear equations will have incorrect data and soon the whole system will be contaminated with nodes having erroneous data.

*Our contribution:* Since the problem of repairing a code is equivalent to wireline network coding [4], existing techniques for network error correction can be used to detect and correct the errors [6], [7]. These techniques are designed to work for general networks and always guarantee a transmission rate of $C - 2z$, where $C$ is the min-cut capacity from the source to the destination and $z$ is the number of links contaminated by the adversary. Our approach, that is creating and communicating small linear hashes which preserve the structure of the code, allows the detection of errors and achieves a transmission rate that can be asymptotically equal to $C$ (by having the receiver connecting to all the non-erroneous nodes) since it takes advantage of the specific structure of the network and the set of links an adversary can contaminate.

To explain our scheme, consider the $(4,2)$ MDS code of Figure 1 and assume one of the four nodes contains errors (say in both rows). A trusted verifier that communicates with all four nodes can find this error by getting the $8$ equations contained in each of the $\binom{4}{2} = 6$ node pairs. Since this is a $(4,2)$ MDS code, the combinations of equations that come from error-free nodes will be full rank and give a consistent solution whereas the other sets will give different solutions (or might not even be full rank). This is, of course, just using the error-correction capability of the code to detect an error. Our contribution involves using this idea to the $N$-extended version of a code, by creating a *linear projection (hash) of each row on the same random vector*. The key observation is that if the same random projection is used, this creates an error-correcting code for the hashes which can be communicated to the verifier. The benefit is that each hash has size only $1/N$ of the data in each row reducing the amount of communication to the verifier. One complication is that each node needs to project its data on the same random vector of length $N$, which requires $N \log q$ bits of common randomness. Subsequently the problem at the verifier is to decode an error-correcting code under adversarial errors. This decoding task can be computationally inefficient but we do not address this issue here, assuming that the verifier can detect the errors if they are within the error correcting capabilities of the code as dictated by the minimum distance (half the minimum distance). Our analysis investigates under which conditions the small projected hash code will detect any error in the large amount of data stored at the nodes. In particular, we prove the following

*Theorem 1:* In a distributed storage system storing a total of $\mathcal{M}$ bits, using an $N$–extended $(n,k)$ MDS code over $\mathbb{F}_q$, with the $n$ storage nodes sharing $O(\mathcal{M})$ bits of common randomness, our random hashing scheme can detect up to $t \leq t_1 \equiv \lfloor (n-k)/2 \rfloor$ errors by communicating a total of $n(n-k)(\log \mathcal{M} + \log t_1)$ bits to a verifier, with probability of failure

$$\mathbb{P}[F] \leq \frac{1}{\mathcal{M}}.$$

One important weakness of the previous result is the large common randomness required which is comparable to the total size of the data object stored ($1/k(n-k)$ fraction of the $\mathcal{M}$ bits). Note that these bits do not have to be a secret, they only need to be realized after the error has been introduced to the new disk. Their large number, however, makes it impractical to generate them at one node and then communicate them to the others. Our second contribution involves showing how to use only $O(\log \mathcal{M})$ bits of common randomness to achieve almost the same performance:

*Theorem 2:* In a distributed storage system storing a total of $\mathcal{M}$ bits, using an $N$–extended $(n,k)$ MDS code over $\mathbb{F}_q$, with the $n$ storage nodes sharing $O(\log \mathcal{M})$ bits of common randomness, our pseudorandom hashing scheme can detect up to $t_1 = \lfloor (n-k)/2 \rfloor$ errors by communicating a total of $O(n(n-k) \log \mathcal{M})$ bits to the verifier, with probability of failure

$$\mathbb{P}[F'] \leq \frac{1}{\mathcal{M}}.$$

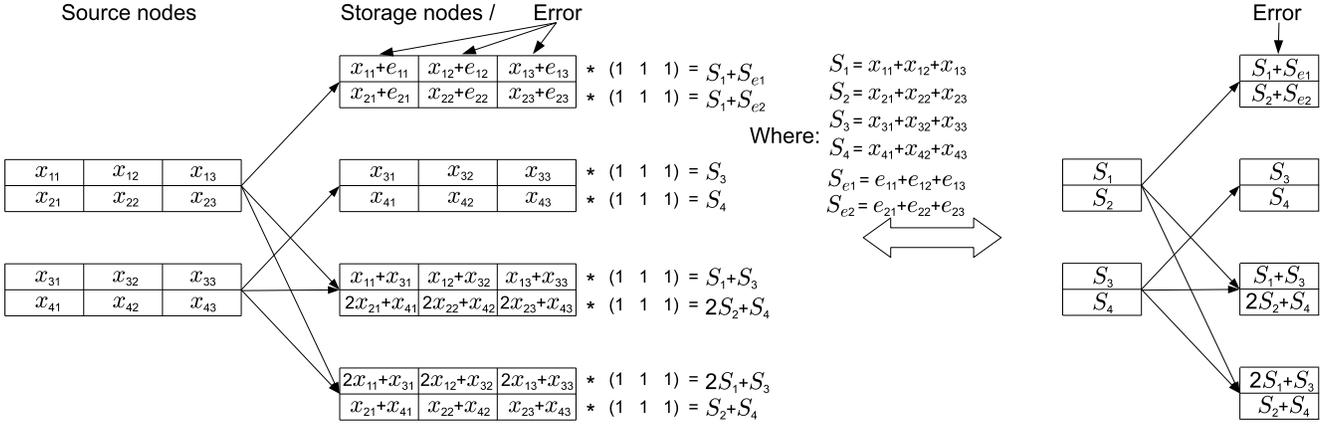

Fig. 2. Illustration of the 3–extended version of the $(4, 2)$ MDS code shown in Figure 1. Each of the three columns stored on the source nodes is coded by repeatedly using the $(4, 2)$ MDS of Figure 1. During verification, each row is projected on the vector $r^T = (1\ 1\ 1)$ and the corresponding products $S_1, \ldots, S_4$ form a codeword of the initial $(4, 2)$ MDS code. For example, the errors at the first row of the first node will not be absorbed by the projection as long as $(e_{11}\ e_{12}\ e_{13}) * (1\ 1\ 1)^T \neq 0$

If there is no common randomness, the verifier can generate the $O(\log \mathcal{M})$ random bits and communicate these to all the nodes requiring a total of $O(n \log \mathcal{M})$ extra communicated bits.

Notice that in this case the total number of bits communicated scales only logarithmically in $\mathcal{M}$, to achieve a probability of failure that scales like $1/\mathcal{M}$. Our construction relies on the pseudorandom small-bias generator used in [8] which can expand $\log N$ random symbols of $\mathbb{F}_q$ (which require $\log N \log q$ random bits to generate), into $N$ pseudorandom symbols that can "fool" any linear function[1]. The only modification to our algorithm is projecting each stored row on this pseudorandom vector to generate each hash and this induces only a small addition to the probability of error. Notice that our work does not rely on any cryptographic assumptions and guarantees that errors inserted in the distributed storage system will be detected with high probability if they are within the capabilities of the code used.

Using the error-correction capability of the code for distributed storage has been suggested before as a way to detect errors [10], [11] and identify "free riders" within the network. A different approach to find errors injected in distributed storage and content distribution systems is the use of signatures and hash functions. Reference [12] introduced the use of homomorphic hashing functions that enables a nodes to perform on-the-fly verification of erasure-encoded blocks. Gkantsidis *et al.* [13] used the computationally less expensive secure random checksums to detect polluted packets in content distribution system that use network coding while [14], [15] used a method of subspace signatures based on different cryptographic primitives. See also [16], [17], [18] for other related work on security and distributed storage.

## II. MODEL

As stated, we consider a data object of size $\mathcal{M}$ bits that is divided into $k$ pieces (of size $\mathcal{M}/k$ bits each) and these are coded into $n\ (>k)$ encoded pieces through a linear $(n,k)$ maximum distance separable (MDS) code. These encoded pieces are stored on $n$ distinct storage nodes along with a header denoting the exact linear combination saved at all the storage nodes. Since the size of the code $(n,k)$ will be much smaller than $N$, the overhead of storing the code description everywhere (including the verifier) is minimal. This simplifies the model and we can now assume that the errors occur only at the data, since an error at the header would be immediately detected.

We assume that the original information (of size $\mathcal{M}$ bits) is organized into a matrix $X$ with $k(n-k)$ rows and $N$ columns. The elements of this matrix are elements of the finite field $\mathbb{F}_q$, *i.e.*, $X \in \mathbb{F}_q^{k(n-k) \times N}$ where $q$ is a prime or an integer power of a prime. Each column $X_i^c \in \mathbb{F}_q^{k(n-k) \times 1}$ ($i \in \{1, \ldots, N\}$) of matrix $X$ will be separately encoded with the use of an $(n,k)$ MDS code with generator matrix $G \in \mathbb{F}_q^{n(n-k) \times k(n-k)}$ and all the columns $GX_i^c \in \mathbb{F}_q^{n(n-k) \times 1}$ derived by this encoding will be stored on the $n$ different storage nodes of the distributed storage system. We will call this code applied to the $N$ different columns of matrix $X$ as the $N$–extended MDS code. The overall effect that the $N$–extended MDS code has upon the information matrix $X$ is captured by the matrix multiplication $GX$. Figure 2 shows such a code for $N = 3$ where the MDS code used is the same as the one shown in Figure 1.

The storage nodes of the distributed storage system are assumed to have limited computational capabilities allowing them only to perform inexpensive operations over the finite field $\mathbb{F}_q$. Some of these storage nodes are assumed to store erroneous information, where these errors might be either random due to hardware failures or inserted adversarially by a malicious user. The malicious user can be computationally unbounded, have knowledge of all the information stored on the distributed storage system and can insert errors to any $t$ of the storage nodes.

We assume the existence of a special node called the *verifier* that is assigned to check the integrity of the data stored on different storage nodes. The verifier does not have access to the initial data object (other than the description of the code) and therefore has to rely on the communicated information to check which nodes contain errors.

## III. RANDOM HASHES

### A. Illustrating example

Assume that in the distributed storage system shown in Figure 2 with four storage nodes it is known that one of them (the first in this example) stores erroneous information. The goal of the verifier that overlooks the state of the whole system is first to find the erroneous disk with the minimum data exchange and second to repair it by using the information stored on the other disks. Since all three columns stored on the distributed storage system are codewords of a $(4,2)$ MDS code with at most one error (some columns might be error free) and minimum distance $d = 3$, the naïve approach to find the erroneous disk is to download all data from different disks and then by using minimum distance decoding on each separate column one would be able to find the erroneous disk.

---
[1]First introduced by Naor and Naor in [9] for linear functions in $\mathbb{F}_2$.

The naïve approach would certainly find the faulty disk but it would require the transfer of double the size of the file stored ($\frac{n}{k}\mathcal{M}$ bits of information in general). So as the size of the file increases this approach will become prohibitively expensive in bandwidth. Instead of transmitting all the information stored on the distributed storage system, the central node could choose a vector with each component chosen independently and uniformly at random from $\mathbb{F}_q$ and have each storage node transmit the inner product (called the hash product) between the randomly chosen vector and each of the rows stored at the disks. In the absence of errors, these hash products will form a codeword of the MDS code used to encode the different columns of the information matrix. In case there are errors, as in the case of the first node in Figure 2, the multiplication with the random vector will not obscure these errors unless $S_{ei} = 0 \Leftrightarrow e_{i1} + e_{i2} + e_{i3} = 0$, for $i = \{1, 2\}$. The reason why the chosen vector should be random is so that the adversary can not deliberately choose the errors to make them "disappear" after the vector multiplication.

*B. General case*

The initial information matrix $X \in \mathbb{F}_q^{n(n-k) \times N}$ is coded with the use of an $N$-extended MDS code with generator matrix $G \in \mathbb{F}_q^{n(n-k) \times k(n-k)}$. Some of the storage nodes contain errors and therefore what is actually stored on the distributed storage system is $Y = GX + E$ where $Y, E \in \mathbb{F}_q^{n(n-k) \times N}$ and $E$ is the error matrix. The verifier wants to identify all erroneous disks by sending hash product requests to all nodes. Then the following theorem holds:

*Proof of Theorem 1:* All storage nodes share $N \log q$ bits of common randomness and therefore they can create the same random vector $r \in \mathbb{F}_q^{N \times 1}$ with each component of vector $r$ drawn uniformly at random from $\mathbb{F}_q$. After the random vector $r$ is computed, each storage node calculates the hash product–inner product–between the random vector $r$ and its content on every row. These $n(n-k)$ hash products are equal to:

$$H = Yr = (GX + E)r \Leftrightarrow H = G(Xr) + e \qquad (1)$$

where $e = Er \in \mathbb{F}_q^{n(n-k) \times 1}$ is a column vector with at most $t_m$ non-zero components representing the erroneous disks (these non-zero components must correspond to the position of at most $t_m$ storage nodes with errors). The key observation is that the projection will not identify an error pattern at a specific row if vector $r$ is orthogonal to that row of $E$. Intuitively, a randomly selected $r$ will be non-orthogonal to an arbitrary row of $E$ with high probability and this is the probability we need to analyze.

From equation (1) it can be seen that the order of applying the MDS encoding on the different columns of the information matrix $X$ and the calculation of the hash products can be interchanged $((GX)r = G(Xr))$ making the process of identifying the erroneous disks equivalent to finding the error positions in a regular MDS code that is guaranteed to succeed if the minimum distance of the code $(n - k + 1)$ is larger than twice the number of errors $2t$ (that is indeed satisfied by the assumptions of Theorem 1).

The set of indices that correspond to the components of vector $e$ that come from disk $i$ is $R_i = \{(i-1)(n-k)+1, \ldots, i(n-k)\}$. We are interested in vector $e$ since this gives us the positions of the faulty disks. One complication that might arise is the fact that disk $i$ might contain an error, meaning that rows $\{E_j^r, j \in R_i\}$ of the error matrix $E$ are not all zero whereas the corresponding components of vector $e$ ($\{e_j, j \in R_i\}$) turn out to be zero and therefore our scheme fails to detect that error. Assume that the set of erroneous disks is $W \subset \{1, 2, \ldots, n\}$ and define $\mathbb{P}[F]$ to be the probability of failing to detect some errors. We get

$$\mathbb{P}[F] = \mathbb{P}\left[\bigcup_{i \in W} \left\{\bigcap_{j \in R_i} (E_j^r \, r = 0)\right\}\right]$$
$$\leq \sum_{i \in W} \mathbb{P}\left[\bigcap_{j \in R_i} (E_j^r \, r = 0)\right] \stackrel{*}{\leq} \sum_{i \in W} \frac{1}{q} \leq \frac{\lfloor \frac{n-k}{2} \rfloor}{q} \equiv \frac{t_1}{q} \quad (2)$$

where inequality $(*)$ holds due to the fact that the probability that some storage node with errors produce zero hash products is less than $1/q^f$ where $f$ is the number of linearly independent errors rows saved at its disk. So by assuming that the adversary has produced linearly dependent errors would only increase the probability of failure.

If the adversary has saved error vectors at storage node $i$ with rank 1 then the probability $\mathbb{P}[\bigcap_{j \in R_i} (E_j^r \, r = 0)]$ in equation (2) reduces to an equation for a single row (assuming row $k$):

$$\mathbb{P}\left[\sum_{e_{kf} \neq 0} e_{kf} r_f = 0\right] = \mathbb{P}\left[r_f = -\sum_{e_{kf'} \neq 0} \frac{e_{kf'}}{e_{kf}} r_{f'}\right] = \frac{1}{q}$$

where we only took the terms with a non-zero error coefficient $e_{kf}$. The numbers $(e_{kf'}/e_{kf}) \, r_{f'}$ ($e_{kf}$ is any non-zero error element from the $k^{\text{th}}$ row) are independent and uniform over $\mathbb{F}_q$ and so is their sum according to Lemma 1. So the last equality holds since two independent uniformly distributed over $\mathbb{F}_q$ random numbers are equal with probability $1/q$.

When the errors have rank $f > 1$ then the probability $\mathbb{P}[\bigcap_{j \in R_i} (E_j^r \, r = 0)]$ can be evaluated by disregarding the linearly dependent rows. By looking only at the linearly independent ones and by choosing $f$ columns we can formulate an invertible submatrix $\hat{E}_i \in \mathbb{F}_q^{f \times f}$ and similarly to the previous analysis we have that $\mathbb{P}[\bigcap_{j \in R_i} (E_j^r \, r = 0)] = \mathbb{P}[\hat{E}_i \hat{r} = \hat{b}]$ where $\hat{r}, \hat{b} \in \mathbb{F}^{f \times 1}$ where $\hat{r}$ are the components of the random vector that correspond to the columns where the submatrix $\hat{E}_i$ was formed. Since $\hat{b}$ is uniformly random, due to the previous analysis $\mathbb{P}[\hat{E} \hat{r} = b] = 1/q^f$.

Each of the $n$ storage nodes has to convey to the verifier the result of the hash product from all its (n-k) rows, so that the total size of the hash communicated is $\mathcal{H} = n(n-k)\log q$, whereas the size of the file $\mathcal{M} = k(n-k)N \log q$. By substituting the field $q$ equal to $\lfloor \frac{n-k}{2} \rfloor \mathcal{M}$ we conclude the proof of Theorem 1. ∎

*Lemma 1:* The sum of any number of independent uniformly distributed random variables gives a uniformly distributed random variable.

*Proof:* Without loss of generality we will prove Lemma 1 only for the case of two random variables. Assume that $x, y \in \mathbb{F}_q$ are two independent and uniformly distributed random variables. We will prove that $x + y$ is also uniformly distributed, indeed $\forall t_1, t_2 \in \mathbb{F}_q$:

$$\mathbb{P}[x + y = t_1] = \sum_{t_2 \in \mathbb{F}_q} \mathbb{P}[x = t_1 - y | y = t_2] \mathbb{P}[y = t_2]$$

$$\stackrel{(*)}{=} \sum_{t_2 \in \mathbb{F}_q} \mathbb{P}[x = t_1 - t_2] \cdot \frac{1}{q} = \sum_{t_2 \in \mathbb{F}_q} \frac{1}{q} \cdot \frac{1}{q} = q \cdot \frac{1}{q^2} = \frac{1}{q}$$

where equality (*) holds due to the independence between $x$ and $y$. ∎

Before we continue to prove Theorem 2 we need to give the following definition (extension of Definition 2.1 in [8] to non-prime numbers):

*Definition 1:* a) Let $q$ be a prime or an integer power of a prime. For a random variable $X$ with values in $\mathbb{F}_q$, let the *bias* of $X$ be defined by

$$\text{bias}(X) = (q-1)\mathbb{P}[X = 0] - \mathbb{P}[X \neq 0]$$

A random variable $X \in \mathbb{F}_q$ is $\epsilon$-*biased* if $|\text{bias}(X) \leq \epsilon|$.

b) The sample space $\mathcal{S} \subseteq \mathbb{F}_q^\ell$ is $\epsilon$-*biased* if for all $c \in \mathbb{F}_q$ and each sequence $\beta = (\beta_1, ..., \beta_\ell) \in \mathbb{F}_q^n \setminus \{0^\ell\}$ the following is valid: if a sequence $X = (x_1, \ldots, x_\ell) \in \mathcal{S}$ is chosen uniformly at random from $\mathcal{S}$, then the random variable $(\sum_{i=1}^{\ell} \beta_i x_i + c)$ is $\epsilon$-biased.

*Proof of Theorem 2:* All storage nodes execute the algorithm described in Proposition 4.1[2] of [8] and produces a pseudorandom vector $r' \in \mathbb{F}_q^{N \times 1}$ with $N$ components. The quantity $m$ in the algorithm (and consequently the field size $\mathbb{F}_{q^m}$ too) is chosen so that the bias $(q-1)(N-1)/q^m$ is equal to 1 and therefore $q^m = (q-1)(N-1)$ or $m = O(\log N)$. The size of the necessary seed that needs to be provided at all the storage nodes so that they can start the algorithm is two elements from $\mathbb{F}_{q^m}$ chosen uniformly at random or equivalently $2m \log q \equiv O(\log N)$ random bits.

Once all storage nodes have constructed the same pseudorandom vector $r'$ they compute the inner product between vector $r'$ and the content stored on each row of the storage nodes. These pseudorandom products are all sent to the verifier to identify the erroneous disks. The whole analysis is identical to the proof of Theorem 1 with one major difference in the calculation of failure probability $\mathbb{P}[F']$. For the case of a pseudorandom vector $r'$, using the same notation as in the proof of Theorem 1:

$$\mathbb{P}[F'] = \mathbb{P}\left[\bigcup_{i \in W} \left\{\bigcap_{j \in R_i} (E_j^r r' = 0)\right\}\right]$$

$$\leq \sum_{i \in W} \mathbb{P}\left[\bigcup_{j \in R_i} (E_j^r r' = 0)\right] \leq \sum_{i \in W} \sum_{j \in R_i} \mathbb{P}(E_j^r r' = 0)$$

$$\stackrel{*}{\leq} (n-k)\lfloor \frac{n-k}{2} \rfloor \frac{2}{q} \equiv \frac{2(n-k)t_1}{q}$$

where inequality (∗) holds since $\mathbb{P}(E_j^r r' = 0) = 2/q$. Indeed the bias of the space constructed by the pseudorandom procedure is 1 that means:

$$|(q-1)\mathbb{P}(E_j^r r' = 0) - \mathbb{P}(E_j^r r' \neq 0)| \leq 1$$

$$\Leftrightarrow |(q-1)\mathbb{P}(E_j^r r' = 0) - [1 - \mathbb{P}(E_j^r r' = 0)]| \leq 1$$

---

[2]This algorithm is described for $q$ prime but it is readily extensible to $q$ equal to an integer power of a prime.

$$\Leftrightarrow \left|q\,\mathbb{P}\left(E_j^r\,r'=0\right)-1\right|\leq 1 \Rightarrow \mathbb{P}\left(E_j^r\,r'=0\right)\leq \frac{2}{q}$$

By setting $q = 2(n-k)t_1\mathcal{M}$ we conclude the proof. ∎

We would like to underline here that both theorems above exhibit the same behavior on the probability. In Theorem 2 the size of the required common randomness is decreased in the expense of an increased field size. Moreover the use of pseudorandom generators incurs the additional computational cost at each storage node of $O(Nm^2)$ or $O(\mathcal{M}\log\mathcal{M})$ operations in $\mathbb{F}_q$ to generate the pseudorandom vector $r'$.

## ACKNOWLEDGMENT

The authors would like to thank Professor Leonard Schulman for his insights on pseudorandom generators.